\documentclass[a4paper]{PoS}
\usepackage{amsmath}
\usepackage{amsthm}
\usepackage{graphicx}
\usepackage{booktabs}

\title{Non-perturbative running of quark masses in three-flavour QCD}
\ShortTitle{Non-perturbative running of quark masses in three-flavour QCD}

\author{%
I.~Campos,$^{a,b}$ %
P.~Fritzsch,$^{b,d}$ %
C.~Pena,$^{b,c}$ %
\speaker{D.~Preti},$^{b}$%
A.~Ramos,$^{d}$ and %
A.~Vladikas$^{e}$%
\vskip0.25em\\
\llap{$^a$} Instituto de F\'isica de Cantabria - IFCA-CSIC, \\
Avda. de Los Castros s/n, 39005 Santander, Spain \\
\llap{$^b$} Instituto de F\'isica Te\'orica UAM/CSIC,
Universidad Aut\'onoma de Madrid, \\
C/ Nicol\'as Cabrera 13-15,
Cantoblanco, Madrid 28049  \\
\llap{$^c$} 
Departamento de F\'isica Te\'orica, Universidad Aut\'onoma de Madrid,\\
Cantoblanco, Madrid 28049   \\
\llap{$^d$}
Theoretical Physics Department, CERN, \\
CH-1211 Geneva 23, Switzerland\\%
\llap{$^e$}
INFN, Sezione di Tor Vergata, c/o Dipartimento di Fisica, Universit\'a di Roma Tor Vergata, \\
Via della Ricerca Scientifica 1, I-00133 Rome, Italy
\vskip0.25em\\
E-mail:~\email{isabel.campos@csic.es}, \email{p.fritzsch@csic.es}, \email{carlos.pena@uam.es}, 
\email{david.preti@csic.es}, \email{alberto.ramos@cern.ch}, \email{vladikas@roma2.infn.it} %
}

\abstract{%
   We present our preliminary results for the computation of the
   non-perturbative running of renormalized quark masses in $N_f=3$ QCD,
   between the electroweak and hadronic scales, using standard finite-size
   scaling techniques. The computation is carried out to very high precision,
   using massless $\mathcal{O}(a)$ improved Wilson quarks. Following the
   strategy adopted by the ALPHA Collaboration for the running coupling,
   different schemes are used above and below a scale $\mu_0 \sim m_b$, which
   differ by using either the Schrödinger Functional or Gradient Flow
   renormalized coupling. We discuss our results for the running in both
   regions, and the procedure to match the two schemes.
}

\FullConference{34th annual International Symposium on Lattice Field Theory\\
24-30 July 2016\\
University of Southampton, UK}

\newcommand{\mNP}{\resizebox{8pt}{!}{NP}}
\begin{document}

\begin{figure}[t!]
  \centering
  \begin{minipage}{0.5\textwidth}
  \centering
  \includegraphics[width=79mm]{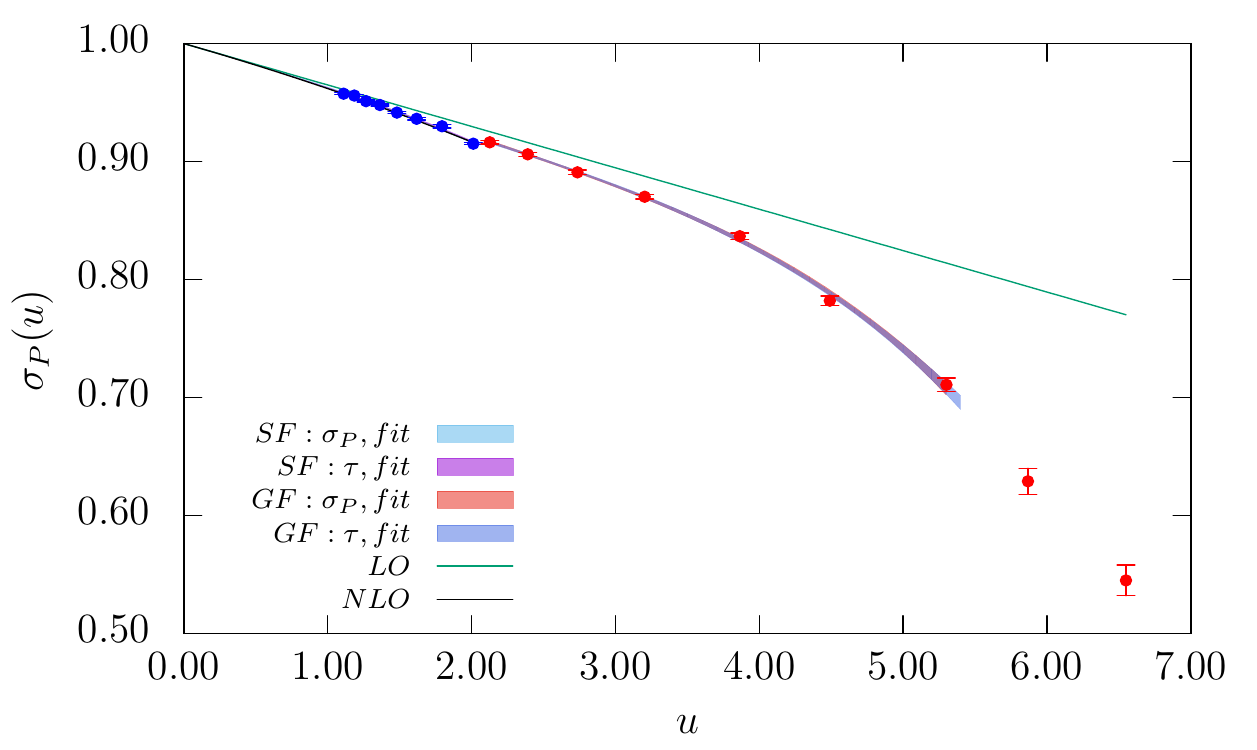}
  \end{minipage}
  \hspace{-1.8mm}
  \begin{minipage}{0.5\textwidth}
  \centering
  \includegraphics[width=79mm]{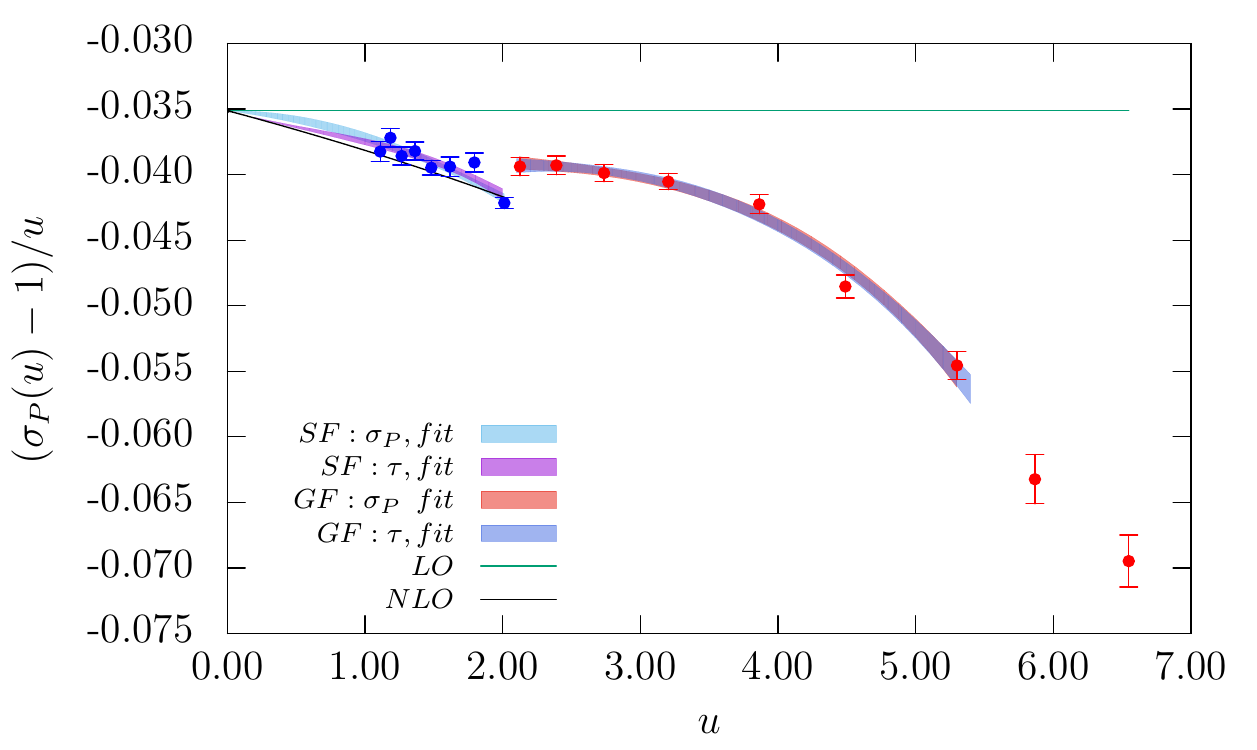}
  \end{minipage}
  \caption{The plots show the SSFs in both SF and GF coupling regions with
           respectively blue and red points (the two most hadronic points have
           not been used in the current analysis). The error bands coming from
           fitting $\sigma_P$ and $\tau$ are laying on top of each others
           showing a remarkable agreement between the two fitting procedures.
           The comparison with the LO and NLO is also provided (the latter is
           known only in the SF coupling region).
          }
  \label{fig:SSF}
\end{figure}

\section{Introduction}

\noindent The high precision computation of quark masses requires to control
the Renormalization Group (RG) running very accurately and in a large range of
scales. The equations describing the RG flow in a mass-independent scheme for
the renormalized coupling $\bar{g}(\mu)$ and the renormalized mass
$\bar{m}(\mu)$ respectively read
\begin{align}
  \mu \frac{\partial}{\partial \mu} \bar{g}(\mu) &=\beta (\bar{g}(\mu))\,,              \label{eq:beta}\\
  \mu \frac{\partial}{\partial \mu} \bar{m}(\mu) &=\tau  (\bar{g}(\mu)) \bar{m}(\mu)\,. \label{eq:tau}
\end{align}
They admit perturbative expansion
\begin{align}
  \beta(g) &\stackrel{g\to0}{\sim} -g^3( b_0 +b_1 g^2 +b_2 g^4 + \mathcal{O}(g^6)) \,,\\
  \tau(g)  &\stackrel{g\to0}{\sim} -g^2( d_0 +d_1 g^2 +d_2 g^4 + \mathcal{O}(g^6)) \,,\label{eq:anomalous_dim}
\end{align}
with universal coefficients $b_0$, $b_1$, $d_0$, while all the others are
scheme-dependent. We can also define through formal solution of
\eqref{eq:beta}, \eqref{eq:tau} the renormalization group invariants (RGI) for
both coupling and mass (the latter is valid for any multiplicatively
renormalizable composite operator \cite{Papinutto:2014xna}) respectively as
\begin{align}
  \Lambda &= \mu \left [ b_0 \bar{g}^2(\mu) \right ]^{-b_1/(2b_0^2)}
             e^{-1/(2b_0\bar{g}^2(\mu))}\exp \left \{ -\int_0^{\bar{g}(\mu)} \, dg \,
             \left [\frac{1}{\beta(g)}+\frac{1}{b_0g^3}-\frac{b_1}{b_0^2g} \right ] \right\} \,,  \\
        M &= \bar{m}(\mu)\left [2b_0\bar{g}^2(\mu) \right ]^{-d_0/(2b_0)}\exp \left \{
            -\int_0^{\bar{g}(\mu)} \, dg \, \left [
            \frac{\tau(g)}{\beta(g)}-\frac{d_0}{b_0g} \right ] \right \} .
\end{align}
In order to compute the running over several orders of magnitude we use a
recursive procedure in finite volume with Schr\"odinger Functional (SF)
\cite{Luscher:1992an} boundary conditions and massless, non-perturbatively
$\mathcal{O}(a)$-improved Wilson fermions. Following the standard SF approach
we identify the scale as the inverse box size $\mu=1/L$ and through a recursive
fine-size scaling $L \to sL$ in the continuum it is possible to compute the
running from large volume simulations ($L \sim 1/\Lambda_{QCD}$) up to the high
energy regions ($L \sim 1/M_W$) where perturbation theory is well defined and
can be safely applied. 

\section{Step Scaling Functions and SF Renormalization Conditions}

\begin{figure}[t]
  \centering
  \includegraphics[width=70mm]{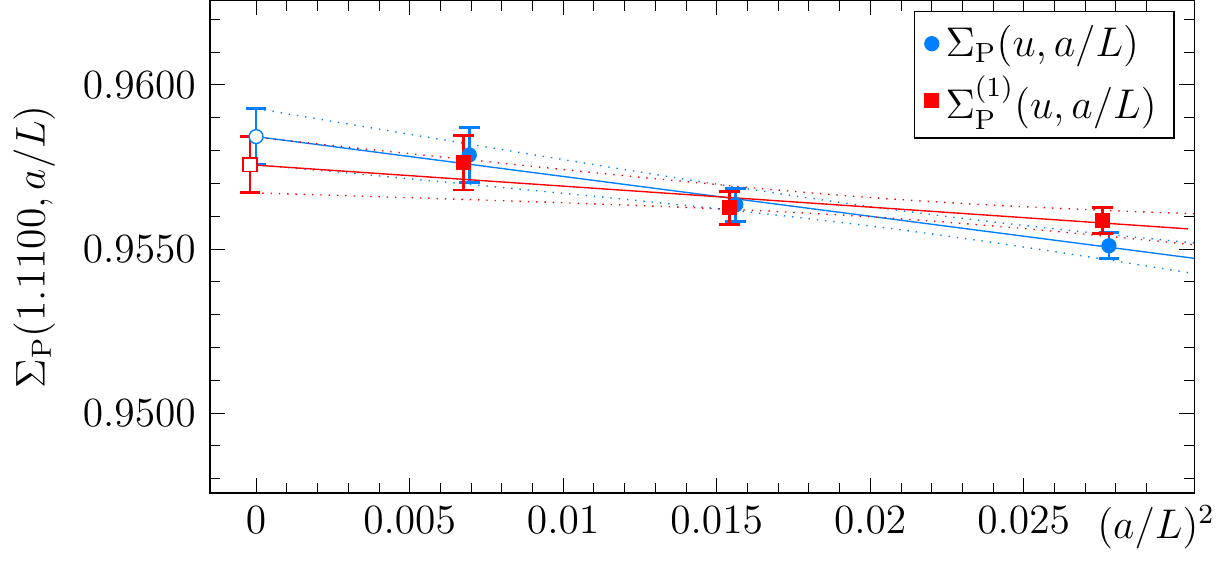}
  \includegraphics[width=70mm]{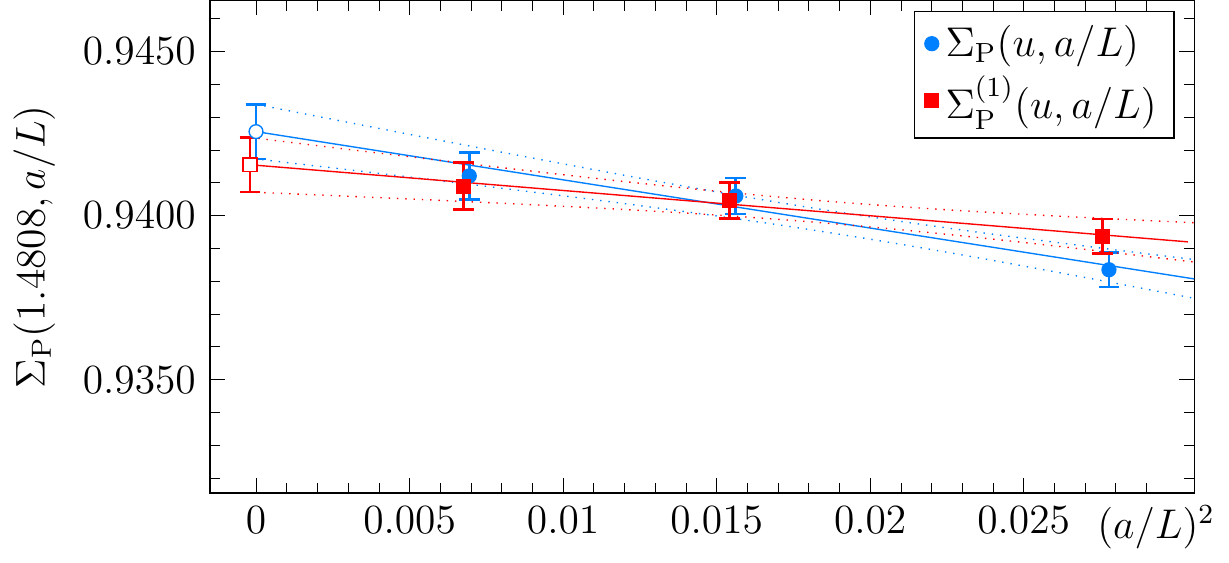}
  \includegraphics[width=70mm]{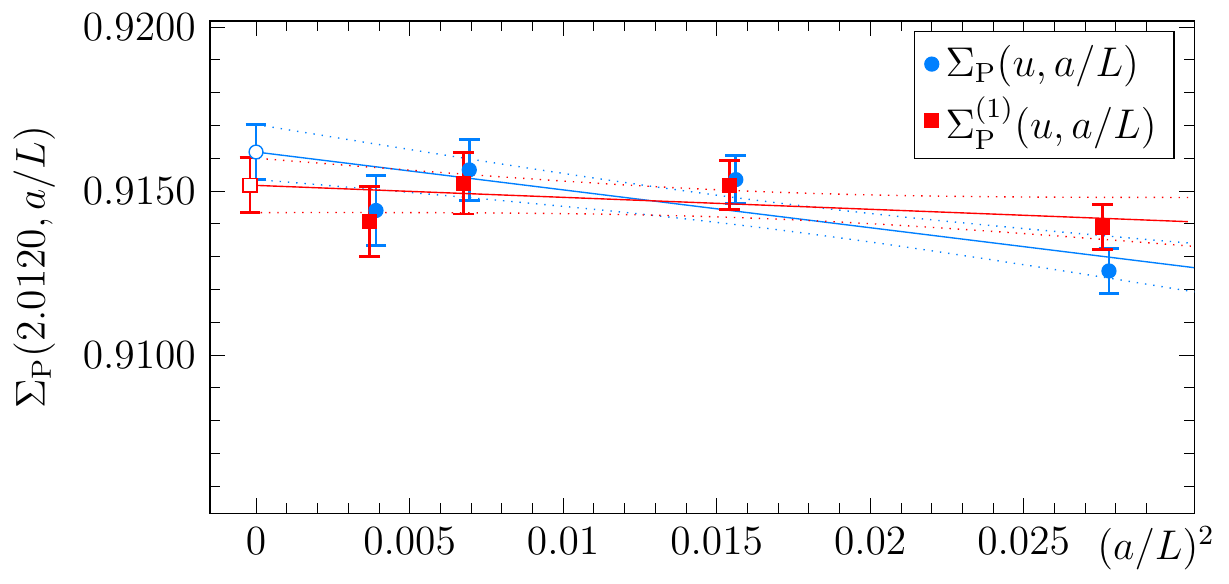}
  \vspace{-4mm}
  \caption{Continuum extrapolations of $\Sigma_P(u)$ for the three values of
           $u_{SF}$ reported in these proceedings. Blue (red) points are raw
           (1-loop improved \cite{Sint:1998iq}) SSFs. The limit $a \to 0$ is
           approached quadratically in $a/L$ according to
           \cite{Luscher:1996sc}.
  }
  \label{fig:Sigma_mat}
\end{figure}

\noindent In our computation the renormalization group functions are accessed
through the Step Scaling Functions (SSFs) $\sigma$ and $\sigma_P$ defining the
scale evolution of a factor $s>1$  for the coupling and the quark mass
respectively as
\begin{align}
  -\ln(s) &= \int_{\sqrt{\bar{g}^2(\mu)}}^{\sqrt{\bar{g}^2(\mu/s)}} \,
             \frac{dg}{\beta(g)} \,, \quad\text{with} \quad
             \sigma(s,\bar{g}^2(\mu))=\bar{g}^2(\mu/s)    \,,          \label{eq:scale_ev_g} \\ 
  \sigma_P(s,\bar{g}^2(\mu)) &= 
             \frac{\bar{m}(\mu)}{\bar{m}(\mu/s)}=\exp  \left \{ - 
             \int_{\sqrt{\bar{g}^2(\mu)}}^{\sqrt{\bar{g}^2(\mu/s)}} \, 
             \frac{\tau(g)}{\beta(g)}dg \right \}         \,.             \label{eq:scale_ev_p}
\end{align}
In order to compute \eqref{eq:scale_ev_p} on the lattice, we identify the
renormalization pattern for the quark masses through the (non-singlet) axial
Ward identity ($i\neq j$)
\begin{align}
   \partial_{\mu}(A_R)^{ij}_{\mu} &= (\bar{m}_i + \bar{m}_j)P^{ij}_R   \,.
\end{align}
The renormalized currents are given by
\begin{align}
   (A_R)^{ij}_{\mu}(x) &= Z_A \bar{\psi}_i(x)\gamma_{\mu}\gamma_5\psi_j(x) \,,  
                          \quad (P_R)^{ij}_{\mu}(x)=Z_P \bar{\psi}_i(x)\gamma_5\psi_j(x) \,,
\end{align}
where $i,j$ are flavour indices. With $Z_A$ being finite, all the scale dependence
of the mass is given by the inverse of the pseudoscalar renormalization
constant $Z_P \propto 1/Z_m$.
The renormalization constant $Z_P$ is computed from standard boundary-to-bulk
and boundary-to-boundary correlation functions in the SF \cite{Capitani:1998mq}
by the renormalization condition 
\begin{align}  \label{eq:ren_cond}
   \left . Z_P(g_0,L/a)\frac{f_P(L/2)}{\sqrt{3f_1}} \right |^{\theta}_{m_q=0} &= c_3(\theta,a/L) \,, &
   \theta &= 0.5 \,,
\end{align}
where $c_3$ is the tree-level normalisation and $\theta$ is entering in the
definition of the boundary quark fields. In particular, the correlation
functions in Eq.~\eqref{eq:ren_cond} are computed with vanishing background
gauge field and quark masses and read 
\begin{align} 
  f_P(x_0) &= -\frac{1}{3}    \int d^3\mathbf{y} \, d^3 \, \mathbf{z} \; 
              \left\langle \bar{\psi}(x)\gamma_5\frac{1}{2}\tau^a\psi(x)\bar{\zeta}(\mathbf{y})\gamma_5\frac{1}{2}\tau^a\zeta(\mathbf{z}) \right\rangle \,, \\
      f_1  &= -\frac{1}{3L^6} \int d^3\mathbf{u}  \, d^3 \mathbf{v} \, d^3 \mathbf{y} \, d^3 \mathbf{z} \; 
              \left\langle \bar{\zeta}'(\mathbf{u})\gamma_5\frac{1}{2}\tau^a\zeta'(\mathbf{v})\bar{\zeta}(\mathbf{y})\gamma_5\frac{1}{2}\tau^a\zeta(\mathbf{z})\right\rangle \,,
\end{align}
using the same notation as in \cite{Luscher:1996sc}. The discrete version of
\eqref{eq:scale_ev_p} for $s=2$ is then given by
\begin{align}
  \Sigma_P(u,a/L) &= \left . \frac{Z_P(g_0,2L/a)}{Z_P(g_0,L/a)} \right |_{u=\bar{g}^2(L)}
\end{align}
from which the continuum limit $\sigma_P(u) = \lim_{a \to 0}
\Sigma_P(u,a/L)$ can be taken. The results for $Z_P$ and $\Sigma_P$
are listed in Tab.~\ref{tab:ZPGF}.  It has been observed 
\cite{deDivitiis:1994yz,DellaMorte:2004bc,Brida:2014joa} that the computational
cost of measuring the SF coupling grows fast at low energies and in particular
towards the continuum limit, thus it is challenging to reach the low energy
domain characteristic for hadronic physics, especially if one aims at
maintaining an high precision. The Gradient Flow (GF) coupling seems to be
better suited for this task
\cite{Fritzsch:2013je,Ramos:2015baa,DallaBrida:2016kgh}. The relative precision
of the coupling in this scheme is typically high and shows a weak dependence on
both the energy scale and the cutoff. Following the same strategy employed by
the ALPHA Collaboration for the computation of the running of the strong
coupling \cite{Brida:2016flw,DallaBrida:2016kgh}, we identify two energy
regions $L>L_0$ and $L<L_0$, where the "switching scale" between the two
schemes $L_{0} \sim 1/m_{b}$ is defined by $\bar{g}^2_{SF}(L_0)=2.012$,
corresponding to $\bar{g}^2_{GF}(2L_0) = 2.6723(64)$. Note that, as part of the
renormalization condition for the mass, the value of the renormalized coupling
is specified.  Therefore, using a different renormalized coupling (e.g. GF or
SF) results in a different renormalization scheme for the mass
\cite{Sint:1998iq}. In the current project we have performed a NP computation
of the SSF for $u_{SF} = [1.1100, 1.1844, 1.2565, 1.3627, 1.4808, 1.6173,
1.7943, 2.0120]$ and $u_{GF} = [2.1257, 2.3900, 2.7359,$ $3.2029, 3.8643,
4.4901, 5.3010]$. In order to compute the continuum extrapolation for the SSF
in both regions we computed the steps $L/a \to 2L/a=[6\to12, 8\to16,12\to24]$
for the SF couplings and $[8\to16,12\to24,16\to32]$ for the GF
couplings\footnote{the step $6 \to 12$ is affected by large cutoff effects
induced by the GF coupling \cite{Fritzsch:2013je,Ramos:2015baa} and it is not
included in the continuum extrapolation}. An exception is given by
$u_{SF}=2.012$ where we added the extra step $16 \to 32$ (as in
Fig.~\ref{fig:Sigma_mat}) in order to have a better control of the continuum
limit since this point plays an important r\^ole in the non-perturbative scheme
matching procedure.

\section{Running and Preliminary results}

\begin{figure}[t]
  \centering
  \includegraphics[width=100mm]{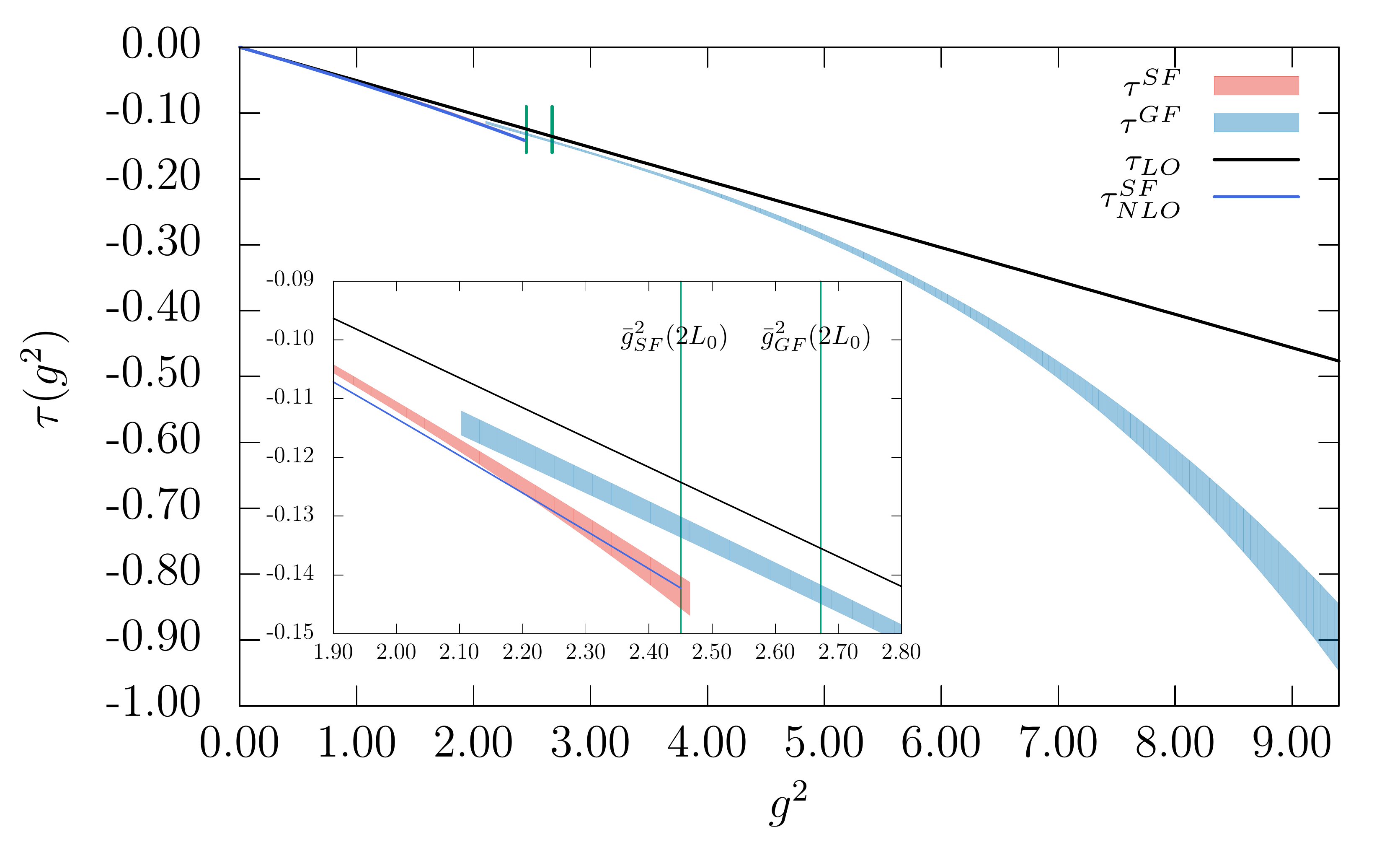}
  \vspace{-4mm}
  \caption{The plot show the comparison among LO and NLO approximation for the
           mass anomalous dimension (while the first is valid for any
           renormalization scheme, the latter is known only in the SF region)
           and our non perturbative computation. The label $\tau^{SF}$ and
           $\tau^{GF}$ correspond to two coupling regions.
          }
  \label{fig:tau}
\end{figure}

\noindent Having the continuum limit $\sigma(u)=\Sigma_P(u,0)$ for various $u$
in both SF and GF regions, to compute the running, we need to obtain an
interpolating function for the SSFs. This can be achieved following two
(equivalent) strategies: perform a polynomial fit with the ansatz
$\sigma_P(u)=1+\bar{p}_0u+p_1u^2 + p_2u^3 + \mathcal{O}(u^4)$ with the first
coefficient fixed to its perturbative value $\bar{p}_0=-d_0\log (2)$, or as an
alternative approach perform a fit directly for the numerator $\tau(g)$ in
\eqref{eq:scale_ev_p}, viz.
\begin{align}
   \sigma_P(s,\bar{g}^2(\mu)) &= 1+\bar{p}_0u+p_1u^2 + p_2u^3 + \mathcal{O}(u^4)
                               = \exp \left \{ - \int_{\sqrt{\bar{g}^2(\mu)}}^{\sqrt{\bar{g}^2(\mu/s)}} 
                               \,\frac{\tau^{\mNP}(g')}{\beta^{\mNP}(g')}dg' \right \} \,.
\end{align}
The anomalous dimension, as depicted in Figure~\ref{fig:tau}, is fitted as a
polynomial 
\begin{align}
  \tau^{\mNP}(g) &=
     \begin{cases}
        -g^2(\bar{d_0}+\bar{d_1}g^2+d_2g^4+d_3g^6 + \mathcal{O}(u^8))  &  \text{when $g\in$ SF region}  \\
        -g^2(d_0+d_1g^2+d_2g^4+d_3g^6 +             \mathcal{O}(u^8))  &  \text{when $g\in$ GF region}
     \end{cases}  \;,
  \label{eq:scale_ev_p_NP}
\end{align}
with coefficients $\bar{d}_i$ fixed to their PT values. The fitted expression
of $\beta^{\mNP}$ in both SF and GF regions
\cite{Brida:2016flw,DallaBrida:2016kgh} is a fundamental input that let us to
isolate the anomalous dimension from the ratio in Eq.~\eqref{eq:scale_ev_p}. As
displayed in Fig.~\ref{fig:SSF} both approaches discussed above completely
agrees within errors. The running from an hadronic scale given by $L_{had}
\sim 200 \, \rm MeV$ can be written as
\begin{align}   \label{eq:running}
   \frac{M}{\bar{m}(L_{had})} &=
       \left . \frac{M}{\bar{m}(L_{pt})} \right |_{SF} \left . \frac{\bar{m}(L_{pt})}{\bar{m}(L_0)} \right |_{SF}  
       \left . \frac{\bar{m}(L_0)}{\bar{m}(2L_0)} \right |_{SF} \left . \frac{\bar{m}(2L_0)}{\bar{m}(L_{had})} \right |_{GF} \,.
\end{align}
The computation of \eqref{eq:running} can be split in the following factors:
the first term on the rhs is the PT matching, computed at NLO in the SF region
with $L_{pt}=L_0/2^N \sim 65 \, \rm GeV$, the second is the standard iterative
procedure carried out with the polynomial interpolation of $\sigma_P(u)$ for
$N=4$ steps (thus gaining a factor $16$ in the scale) given by
\begin{align}\label{eq:2}
     \left . \frac{\bar{m}(L_{pt})}{\bar{m}(L_0)} \right |_{SF} &= \prod_{i=1}^N \sigma_P(u_i) \,,  &&\text{with}  &
                                                \sigma(u_{i+1}) &= u_i = \bar{g}^2(2^{-i}L_0) \,, 
\end{align}
the third factor (that could be included in the iteration above) represents the
NP scheme matching since it is connecting the two coupling regions 
\begin{align}\label{eq:3}
    \left . \frac{\bar{m}(L_0)}{\bar{m}(2L_0)} \right |_{SF} &= \sigma_P(u_0) \,, &&\text{with} & 
                                                         u_0 &= \bar{g}^2(L_0)=2.012 \,.
\end{align}
The last term is then the ratio of the running from an hadronic scale to the
scheme-switching scale. In order to have more flexibility in choosing $L_{had}$
we take advantage of the non-perturbative RG functions and directly determine  
\begin{align}
  \left . \frac{\bar{m}(2L_0)}{\bar{m}(L_{had})} \right |_{GF} &= 
    \exp \left \{ -\int_{\bar g(2L_0)}^{\bar g(L_{had})}  \frac{\tau_{GF}^{\mNP}(g)}{\beta_{GF}^{\mNP}(g)}\, dg \right \} \,.
\end{align}
Note that here we are computing an integral whose limits are the two scales we
want to connect by RG evolution. They do not have to be any more related by an
integer scaling factor $s$, as it is being applied in the SSF recursion. In the
present work, identifying the hadronic scale with the one corresponding to the
most hadronic point%
\footnote{note that this is not the final value defining $L_{had}$} 
that is covered by the SSF from $u_{GF}=5.3010$, we obtain
$L_{had}/L_0=18.74(26)$. With this choice, the total range of scales covered by
the non-perturbative running in both schemes is $L_{had}/L_{pt}=300(4)$ and
finally the total RG running factor $M/\bar{m}(L_{had})=0.9088(78)$.  

\begin{table}
   \centering
  \begin{tabular}{ccccccc}
  \toprule
  $u_{\rm SF}$ & $L/a$ & $\beta$ & $\kappa$ & $Z_P(g_0^2,L/a)$ & $Z_P(g_0^2,2L/a)$ & $\Sigma_P(u,a/L)$ \\
  \midrule
         & 6   &  8.5403 & 0.13233610 & 0.80494(22) & 0.76879(24) &  0.95510(40)   \\
  1.1100 & 8   &  8.7325 & 0.13213380 & 0.79640(22) & 0.76163(34) &  0.95635(50)   \\
         & 12  &  8.9950 & 0.13186210 & 0.78473(29) & 0.75167(59) &  0.95786(83)   \\ \midrule
         & 6   &  7.2618 & 0.13393370 & 0.75460(27) & 0.70808(31) &  0.93835(53)   \\
  1.4808 & 8   &  7.4424 & 0.13367450 & 0.74425(26) & 0.70004(33) &  0.94060(55)   \\
         & 12  &  7.7299 & 0.13326353 & 0.73515(33) & 0.69193(42) &  0.94121(72)   \\ \midrule
         & 6   &  6.2735 & 0.13557130 & 0.69013(32) & 0.62979(37) &  0.91256(68)   \\
  2.0120 & 8   &  6.4680 & 0.13523620 & 0.68107(28) & 0.62341(43) &  0.91535(74)   \\
         & 12  &  6.72995& 0.13475973 & 0.67113(43) & 0.61452(49) &  0.91564(93)   \\
         & 16  & 6.93460 & 0.13441209 & 0.66627(31) & 0.60924(66) &  0.9144(11)    \\ \toprule
  $u_{\rm GF}$ & $L/a$ & $\beta$ & $\kappa$ & $Z_P(g_0^2,L/a)$ & $Z_P(g_0^2,2L/a)$ & $\Sigma_P(u,a/L)$ \\ \midrule
         & 8   &  5.3715 & 0.13362120 & 0.73275(27) & 0.67666(64) &  0.9234(9)     \\
  2.1257 & 12  &  5.5431 & 0.13331407 & 0.71301(32) & 0.65750(89) &  0.9221(13)    \\
         & 16  &  5.7000 & 0.13304840 & 0.70248(32) & 0.64369(86) &  0.9163(13)    \\ \midrule
         & 8   &  4.4576 & 0.13560675 & 0.64779(33) & 0.56891(75) &  0.8782(12)    \\
  3.2029 & 12  &  4.6347 & 0.13519986 & 0.62622(42) & 0.54749(94) &  0.8743(16)    \\
         & 16  &  4.8000 & 0.13482139 & 0.61735(46) & 0.5382(11)  &  0.8718(19)    \\ \midrule
         & 8   &  3.7549 & 0.13701929 & 0.52174(47) & 0.3924(29)  &  0.7522(55)    \\
  5.3010 & 12  &  3.9368 & 0.13679805 & 0.50366(53) & 0.3652(21)  &  0.7251(42)    \\
         & 16  &  4.1000 & 0.13647301 & 0.49847(73) & 0.3609(23)  &  0.7240(48)    \\
  \bottomrule
  \end{tabular}
  \caption{Example of results for $Z_P$, $\Sigma_P$ in both $\rm SF$ and 
           $\rm GF$ region.
          }
  \label{tab:ZPGF}
\end{table}

\section{Conclusions}

\noindent We have computed the NP running quark mass in three-flavour QCD
between $\sim 200 \, \rm MeV$ and $\sim 60\,  \rm GeV$ with an unprecedented
sub-percent uncertainty. This is a major achievement compared to a similar
determinations in the past~\cite{DellaMorte:2004bc,Aoki:2010wm}. In order to
optimise the overall precision (in particular towards the hadronic scales), we
have employed two different schemes equipped with a non-perturbative at an
intermediate scale of $\sim 4 \, \rm GeV$. 
Another completely new results is given by the computation of the NP
mass anomalous dimension for both SF and GF coupling regions allowing for a
more flexible choice of the hadronic matching scale. 

\section*{Acknowledgments}

\noindent The simulations were performed on the Altamira HPC facility, the
GALILEO supercomputer at CINECA (INFN agreement), Finisterrae-2 at CESGA and
CERN. We thankfully acknowledge the computer resources and technical support
provided by the University of Cantabria at IFCA, CESGA, CINECA and CERN.
P.F. acknowledges financial support from the Spanish MINECO's ``Centro de
Excelencia Severo Ochoa'' Programme under grant SEV-2012-0249, as well
as from the grant FPA2015-68541-P (MINECO/FEDER). 

\bibliography{mainbib}

\begin{thebibliography}{10}

\bibitem{Papinutto:2014xna}
M. Papinutto, C. Pena and D. Preti,
\newblock PoS LATTICE2014 (2014) 281,
  [\href{http://xxx.lanl.gov/abs/1412.1742}{{1412.1742}}].

\bibitem{Luscher:1992an}
M. L{\"u}scher et~al.,
\newblock Nucl.Phys. B384 (1992) 168,
  [\href{http://xxx.lanl.gov/abs/hep-lat/9207009}{{hep-lat/9207009}}].

\bibitem{Sint:1998iq}
ALPHA, S. Sint and P. Weisz,
\newblock Nucl.Phys. B545 (1999) 529,
  [\href{http://xxx.lanl.gov/abs/hep-lat/9808013}{{hep-lat/9808013}}].

\bibitem{Luscher:1996sc}
M. L{\"u}scher et~al.,
\newblock Nucl.Phys. B478 (1996) 365,
  [\href{http://xxx.lanl.gov/abs/hep-lat/9605038}{{hep-lat/9605038}}].

\bibitem{Capitani:1998mq}
ALPHA, S. Capitani et~al.,
\newblock Nucl.Phys. B544 (1999) 669,
  [\href{http://xxx.lanl.gov/abs/hep-lat/9810063}{{hep-lat/9810063}}].

\bibitem{deDivitiis:1994yz}
ALPHA, G. de~Divitiis et~al.,
\newblock Nucl.Phys. B437 (1995) 447,
  [\href{http://xxx.lanl.gov/abs/hep-lat/9411017}{{hep-lat/9411017}}].

\bibitem{DellaMorte:2004bc}
ALPHA, M. Della~Morte et~al.,
\newblock Nucl.Phys. B713 (2005) 378,
  [\href{http://xxx.lanl.gov/abs/hep-lat/0411025}{{hep-lat/0411025}}].

\bibitem{Brida:2014joa}
ALPHA, P. Fritzsch et~al.,
\newblock PoS LATTICE2014 (2014) 291,
  [\href{http://xxx.lanl.gov/abs/1411.7648}{{1411.7648}}].

\bibitem{Fritzsch:2013je}
P. Fritzsch and A. Ramos,
\newblock JHEP 1310 (2013) 008,
  [\href{http://xxx.lanl.gov/abs/1301.4388}{{1301.4388}}].

\bibitem{Ramos:2015baa}
A. Ramos and S. Sint,
\newblock Eur. Phys. J. C76 (2016) 15,
  [\href{http://xxx.lanl.gov/abs/1508.05552}{{1508.05552}}].

\bibitem{DallaBrida:2016kgh}
ALPHA, M. Dalla~Brida et~al.,
\newblock (2016), [\href{http://xxx.lanl.gov/abs/1607.06423}{{1607.06423}}].

\bibitem{Brida:2016flw}
ALPHA, M. Dalla~Brida et~al.,
\newblock Phys. Rev. Lett. 117 (2016),
  [\href{http://xxx.lanl.gov/abs/1604.06193}{{1604.06193}}].

\bibitem{Aoki:2010wm}
PACS-CS, S. Aoki et~al.,
\newblock JHEP 1008 (2010) 101,
  [\href{http://xxx.lanl.gov/abs/1006.1164}{{1006.1164}}].

\end{thebibliography}

\end{document}